\documentclass[a4paper,11pt]{amsart}

\usepackage{hyperref}
\hypersetup{colorlinks=true,linkcolor=black}
\usepackage[english]{babel}
\usepackage{graphicx}

\usepackage[small,bf]{caption}   

\title[Return Radius and  volume of recrystallized material]{Return Radius and  volume of recrystallized material in Ostwald Ripening}

\author[F. Hau{\ss}er] {Frank Hau{\ss}er}
\address{Frank Hau{\ss}er\\
        Beuth Hochschule Berlin, University of Applies Sciences \\
        Luxemburger Stra{\ss}er 10\\
        13353 Berlin, Germany}
\email{hausser@beuth-hochschule.de}

\author[E. Laksthanov]{Evgeny Lakshtanov}
\address{Evgeny Lakshtanov \\
Department of Mathematics, Aveiro University \\
Aveiro 3810, Portugal}
\email{lakshtanov@rambler.ru}

\date{July 6th, 2012}


\newcommand{\new}[1]{{\color{black} #1}}

\begin{document}

\begin{abstract}
Within the framework of the LSW theory of Ostwald
ripening the amount of volume of the second (solid) phase \new{in a liquid solution}
that is newly formed  by recrystallization is investigated.
It is shown, that in the late stage, the portion of the
newly generated volume formed within an interval from time $t_0$
to $t$ is a certain function of $t/t_0$ and an explicit expression of this volume is given.
To achieve this, we introduce the notion of the  {\it return radius} $r(t,t_0)$, which is the unique radius
of a particle at time $t_0$ such that this particle has -- after growing and shrinking -- the same radius at time $t$.
We derive a formula for the return radius which later on is used to obtain the newly formed volume.
Moreover, formulas for the growth rate of the return radius and the recrystallized material at time $t_0$ are derived.
\end{abstract}

\maketitle


\section{Introduction}
Recrystallization of minerals is a combination of simultaneous processes of dissolution and precipitation that leads to formation of larger mineral crystals. A driving force of recrystallization in geological environments is usually either the difference between lithostatic and hydrostatic pressures or the dependence of the chemical potential of the interface on the grain size. This type of coarsening process, where larger particles are growing at the expense of smaller ones is called Ostwald ripening. A prominent example is Ostwald ripening of  calcite
($\mathrm{CaCO_3}$)  in an aqueous solution, which is subject of a number of experimental investigations, see e.g.
 \cite{InksHahn1967},\cite{Mozeto1984}, \cite{Davis1987},\cite{Zachara1991},\cite{Curti2005},\cite{Belova2012}.

In these experiments one is interested in the amount of newly formed crystalline material, i.e. the volume of the solid phase present at time $t$, that has been precipitated from the solution after some time instant $t_0 < t$.
Experimental methods \new{tending} to determine the newly formed material are usually proceeding as follows: An isotope of a crystal lattice constituent of the mineral -- in case of calcite $\mathrm{^{45}C\!a}$ or $\mathrm{^{14}C}$ -- is added as  a radioactive tracer to the solution at time $t_0$.
The isotope concentration in the solution is monitored during the experiment, and changes of the isotope concentration happen due to two processes: either it is a diffusion into the crystal, or burial in the calcite layers  newly formed during recrystallization.

In order to derive the amount of newly formed material from the Isotope uptake,
in \cite{Belova2012},\cite{Davis1987},\cite{Curti2005} it was argued that the amount of isotope uptake is proportional to the newly formed material. However, this approach does not account for the surface changes during ripening. Whenever the surface area decreases, adsorbed isotope will be returned to the solution which does influence the overall isotope uptake. The question we were asked by experimentalists was if, and under which conditions, the surface change may be neglected or, more precisely, how the influence of surface change compares with the amount of newly formed material concerning the isotope uptake. To answer these questions, a first step is to derive analytical expressions both, for the newly formed volume and for the change of the surface area. The former will be the subject of this paper, while the latter usually may be derived very easily in some mean field theory.
We will not attempt to draw conclusions concerning experiments, since this would require too many details of geochemistry in order to construct a sound mathematical model. This is planned to be done in a forthcoming article. Besides these motivations from experiments, an explicit formula for the newly formed volume during Ostwald Ripening is also interesting from a theoretical point of view in order to better understand the long time behavior of the coarsening system.

\begin{figure}[htbp]
\begin{center}	
	\includegraphics*[width=0.5\textwidth]{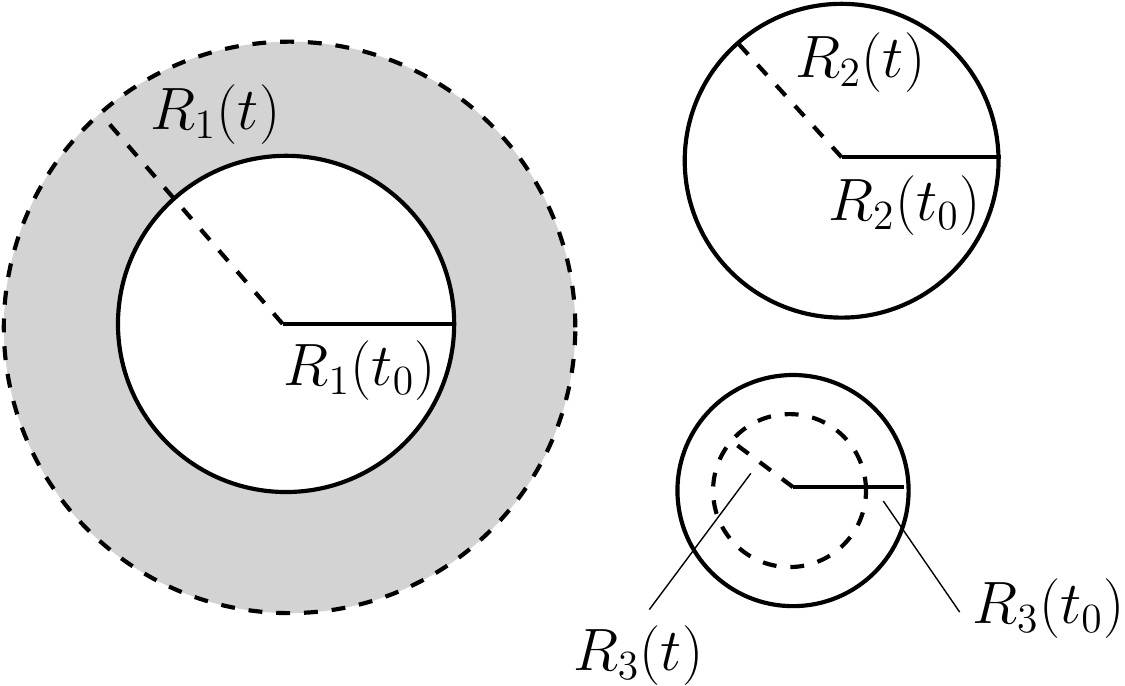}
	\caption{Newly formed solid phase.
	During Ostwald ripening, bigger particles are growing at the expense of smaller ones. Comparing the sizes at time $t_0$ and at time $t>t_0$, the three possible cases are sketched: $R_1(t) > R_1(t_0)$,  $R_2(t) = R_2(t_0)$ and $R_3(t) < R_3(t_0)$.
  Thus at time	$t > t_0$,  a certain amount of the solid phase, marked in gray, has been newly formed between time $t_0$ and time $t$.}
	\label{fig:ripening}
\end{center}
\end{figure}

\begin{figure}[htbp]
\begin{center}	
	\includegraphics*[width=0.55\textwidth]{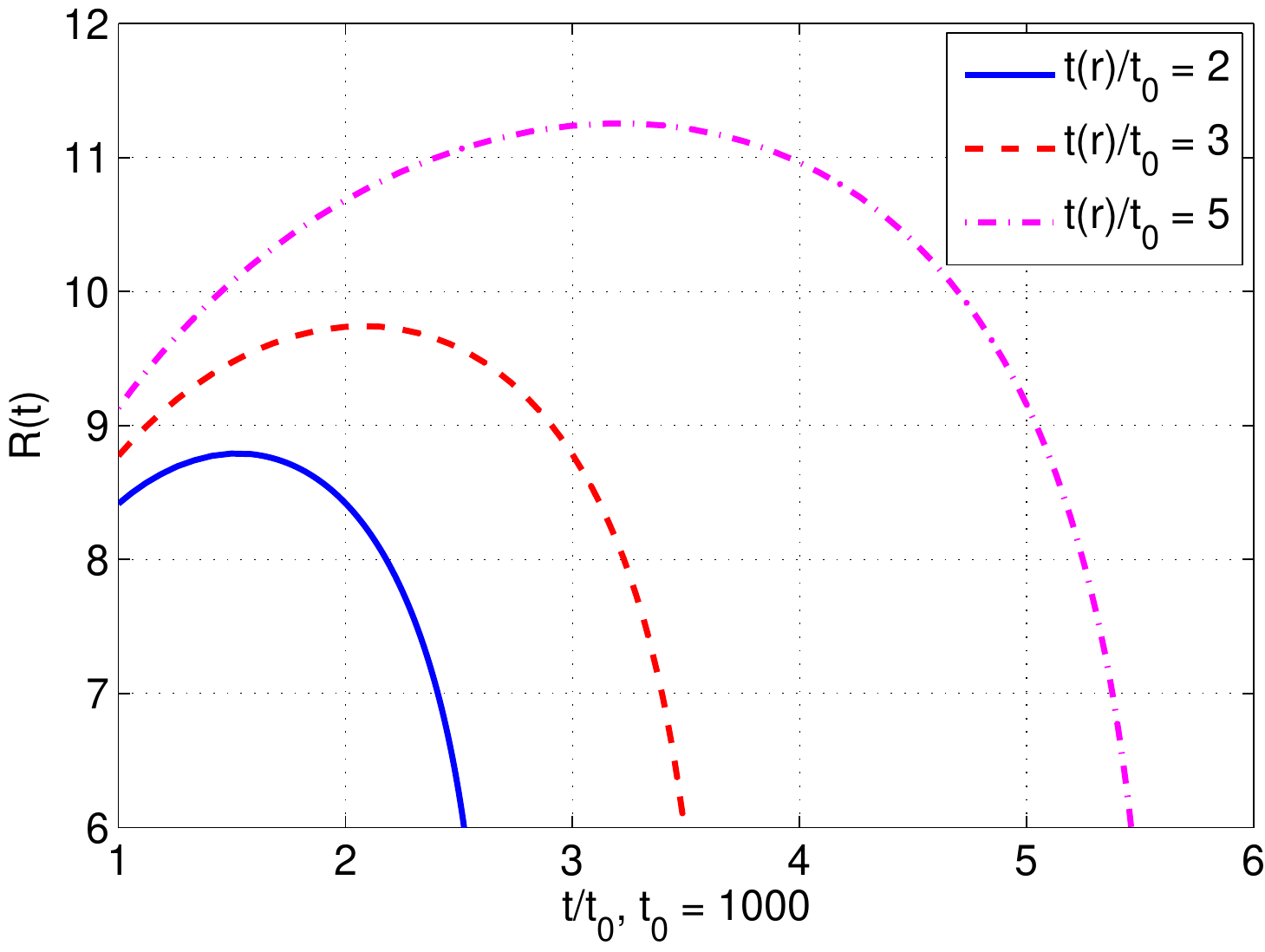}
	\caption{Time evolution of the radius $R(t)$ of a spherical particle with different initial conditions $R_0 = R(t_0)$ in the late stage of Ostwald ripening in the diffusion limited LSW regime as described in section~\ref{s:LSW}, with non dimensional critical radius $R_c(0) = 1$.
	Since in all three cases the initial radius $R(t_0)$ is larger than the critical radius $R_c(t_0)$, the particles start to grow at the expense of smaller particles. However, after a certain time  they start shrinking at the expense of even bigger particles, since the critical radius is also growing in time.
For a particle radius $R(t_0) = R_0$ being larger than the critical radius at time $t_0$, there is a {\it return time} $t(R_0)$ such that $R(t) = R_0$. The other way round, given $t > t_0$, there is a unique radius $r$	such that a particle with initial radius $R(t_0) = r$ will have the same radius at time $t$, i.e. $R(t) = r$. This radius will be called the {\it return radius}.  E.g. for the blue line we observe that $R(2t_0) = R(t_0)$, i.e., the return time is $t=2t_0$ and the return radius for $t = 2t_0$ is given by $R(t_0)\approx 8.42$. In fact, we have calculated suitable initial values $R(t_0)$ for the above cases with our method described in section~\ref{s:radius}.}
	\label{fig:radiusOverTimeDL}
	\end{center}
\end{figure}

In this paper we are going to calculate the newly formed volume theoretically in the late stage of the coarsening process, assuming that the so called  LSW theory is valid, see \cite{LifshitzSlyozov1961}, \cite{Wagner1961}. In this mean field theory, it is assumed, that the grains are spherical particles and the growth kinetics of a grain only depends on its size compared to the size distribution of all particles and not on the local environment. \new{Moreover it is assumed, that the total volume of the crystallized material is conserved, i.e. large particles grow exclusively at the expense of smaller ones.}. A short review of the LSW theory is given in the next section.
In Figure~\ref{fig:ripening} we have sketched how the size of particles may have changed from time $t_0$ to $t$. Note that a particle may grow at time $t_0$ but start to shrink at some $t_1$ with $t_0 < t_1 < t$ leading to $R(t) < R(t_0)$, see Figure~\ref{fig:radiusOverTimeDL}, where some typical examples of the time evolution $R(t)$ of particle radii $R$, are shown.

For an ensemble of particles with radii $R_i(t_0)$, the newly formed volume  $V^\mathrm{new} (t,t_0)$ between $t_0$ and $t$ is	
	\begin{equation}
	\label{eq:intro1}
	V^\mathrm{new} (t,t_0) =  \tfrac{4}{3} \pi \sum_{i: R_i(t) \geq R_i(t_0)} \big(R_i(t)^3 - R_i(t_0)^3 \big).
	\end{equation}
	\new{Note that due to the assumption of mass conservation of the second phase, the newly formed crystalline material all comes from dissolution of smaller particles.}
	In a mean field picture $R_i(t) \geq R_j(t)$ if and only if $R_i(t_0) \geq R_j(t_0)$.
	\new{As will be discussed in more details in the next sections, there is a critical radius $R_c(t)$ such that at time $t$ precisely the particles with radius $R(t) > R_c(t)$ are growing. However, $R_c(t)$ is growing faster then $R(t)$ and particles that initially grow, start to shrink at later times, see Figure~\ref{fig:radiusOverTimeDL}.}
Thus, there is a unique radius $r = r(t,t_0)$, such that
	$R_i(t_0) = R_i(t)$ if and only if $R_i(t_0) = r$. We will call this radius $r$ the {\it return radius} and $V^\mathrm{new} (t,t_0)$
	may be expressed as
	\begin{equation}
	\label{eq:intro2}
	V^\mathrm{new} (t,t_0) =  \tfrac{4}{3} \pi \sum_{i: R_i(t) \geq r(t,t_0)} \big(R_i(t)^3 - R_i(t_0)^3 \big).
	\end{equation}	

We will show in section~\ref{s:radius} that in the late stage of Ostwald ripening in the LSW regime the return radius $r(t,t_0)$ is a function of $t/t_0$ and can be easily computed by inverting an explicitly given function. Moreover we obtain an analytic expression for the growth rate of $r(t,t_0)$ at $t = t_0$. This will lead in section~\ref{s:volume} to our main result: Also the volume 	$V^\mathrm{new} (t,t_0)$ or equivalently the volume fraction $\Phi^\mathrm{new}(t,t_0)$ of the newly formed solid phase within the interval from $t_0$ to $t$ depends on $t/t_0$ and may be calculated by function inversion of explicitly given expressions. We also give an explicit expression for the initial rate of the formation of new solid material.


\section{LSW Theory for Ostwald Ripening}
\label{s:LSW}
We will shortly review the main results of the LSW-analysis \new{given in \cite{LifshitzSlyozov1961},\cite{Wagner1961}}, for a more detailed description see e.g. \cite{RadtkeVoorhees2002}.
\new{In the following the three dimensional model is presented}, using the notations of \cite{HausserVoigt2005e}, where the two dimensional case is discussed.
The  kinetics of Ostwald Ripening is governed by  two different processes: the mass transport between the mineral grains via diffusion in the solution and the attachment/detachment process at the grain boundaries. The two limiting kinetic regimes are termed {\it attachment limited} (AL) growth -- here diffusion is assumed to be infinitely fast compared to the attachment process -- and {\it diffusion limited} (DL) growth -- assuming the attachment process to be instantaneous.

The first assumption of LSW-theory is, that the volume fraction $\Phi$ of the dispersed solid phase is very small. Thus it is assumed that the solid phase consists of many disconnected particles far away from each other which moreover are assumed to be spherical and have immobile center. For a large system, this ensemble of particles
may be characterized in terms of a
particle radius distribution function $F(R,t)$. The number of particles per unit volume is then given by $n(t) = \int_0^\infty
F(R,t)\, dR $. Assuming that no nucleation and coalescence of particles takes place, $F$ obeys the continuity
equation
\begin{equation}
  \label{eq:LSW1}
  \partial_t F + \partial_R \big( \dot{R} F \big) = 0.
\end{equation}
Note that the number of particles $n(t)$ may only change due to radii decreasing to zero in finite time.

\begin{figure}[htbp]
\begin{center}	
	\includegraphics*[width=0.55\textwidth]{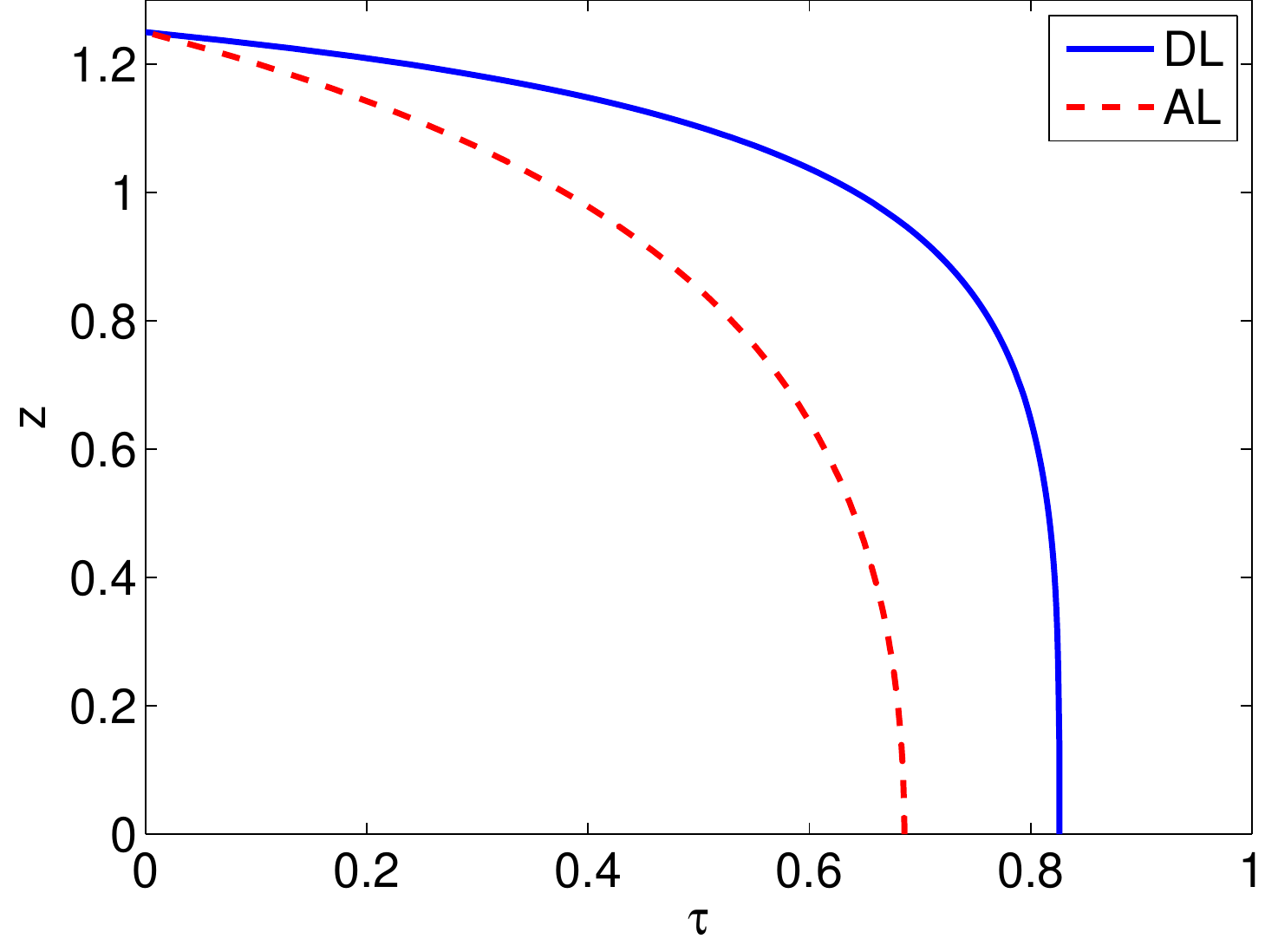}
	\caption{The rescaled radius $z(\tau)$ for a particle with initial value $z(0) = 5/4$ obtained by inverting $\tau(z)$ given in \eqref{eq:tau(z)}. Note that despite the fact, that the initial particle radius is larger than the critical rescaled radius $z=1$, the rescaled radius $z(\tau(t)) = R(t)/R_c(t)$ is decreasing in time.}
	\label{fig:zTau}
	\end{center}
\end{figure}

The second assumption is, that far away from the particles, the chemical potential $u(x,t)$, \new{describing the change of energy per change of mass of the dissolved material} may be approximated by a spatially constant mean field $\bar{u}(t)$.
Finally, one assumes that the volume fraction $\Phi$ is constant in time. \new{The driving force of Ostwald ripening is the difference between the chemical potential  $\bar{u}(t)$ in the solution and the chemical
$u_\mathrm{eq}(R)$ at the particle surface, depending on the curvature $1/R$.}
From the above assumptions it is derived, that the growth rate $\dot{R}(t)$ of any particle may be described in non-dimensional form as
\begin{align*}
	\dot{R}(t)& = \big( \bar{u}(t) - u_\mathrm{eq}(R) \big) \frac{1}{R}, && \quad
	   \bar{u}(t) =  \frac{\sum u_\mathrm{eq}(R_i) R_i}{\sum R_i}    && \quad \text{(DL)} \\
	\dot{R}(t)& = \big( \bar{u}(t) - u_\mathrm{eq}(R) \big), &&  \quad
	  \bar{u}(t) =  \frac{\sum u_\mathrm{eq}(R_i) R_i^2}{\sum R_i ^2} && \quad \text{(AL)}.
\end{align*}
\new{Moreover, assuming the chemical potential at the surface to be given in non dimensional form by the Gibbs Thomson law $u_\mathrm{eq}(R) = 1/R$, the critical radius $R_c$ such that particles with radius $R < R_c$ are shrinking while the ones with $R > R_c$ are growing} is given by
\[
R_c(t) = 1/\bar{u}(t)
\]
and the growth law may be written as
\begin{align}
\label{eq:vel_DL}
	\dot{R}(t)& =  \frac{1}{R^2}\Big( \frac{R}{R_c} - 1 \Big), \quad
	   \bar{u}(t) =  \frac{1}{\overline{R}} && \quad \text{(DL)}\\
	   \label{eq:vel_AL}
	\dot{R}(t)& =  \frac{1}{R}\Big( \frac{R}{R_c} - 1 \Big), \quad
	   \bar{u}(t) =  \frac{\overline{R}}{\overline{R^2}}  && \quad \text{(AL)}.
\end{align}
Introducing the new variables
\begin{equation}
  \label{eq:resc}
  z = \frac{R}{R_c}, \qquad  \tau = \ln(\frac{R_c(t)}{R_c(0)}),
\end{equation}
eqs.~\eqref{eq:vel_DL},\eqref{eq:vel_AL} become
\begin{gather}
  \label{eq:vel_rescaled}
 \frac{d z}{d\tau} = \nu\,\frac{z - 1}{z^\lambda} - z,
\end{gather}
where $\lambda = 2$ (DL) or $\lambda = 1$ (AL) and $\nu$ is a function of the critical radius $R_c$:
\begin{equation}
  \label{eq:nudk}
  \nu = \frac{1}{R^2_c \, \dot{R}_c(t)}, \quad \text{(DL)} \qquad
  \nu =  \frac{1}{R_c \, \dot{R}_c(t)},  \quad \text{(AL)}.
\end{equation}
Note that in the new coordinates, $z = 1$ corresponds to the critical radius $R_c(t)$ of a particle, that is neither growing nor shrinking. \new{For later use we also point out, that according to \eqref{eq:vel_rescaled} the rescaled radius $z$ may be shrinking also for values $z>1$, i.e. for growing particles. Actually, as will be seen later, the growth rate $dz/d\tau$ is negative for all particles in the late stage.}
The essential point of the LSW-analysis is, to argue,  that $\nu$
 becomes constant at late times approaching the unique values \new{(\cite{Wagner1961},\cite{LifshitzSlyozov1961})}
 \begin{equation}
 \label{eq:nu}
 \nu = 27/4 \quad \text{(DL)}, \qquad  \nu = 4 \quad \text{(AL)}.
 \end{equation}
 This in turn implies a  scaling law for the critical radius $R_c$ by solving the differential equations for $R_c(t)$ given in Eq.~\eqref{eq:nudk}. \new{Assuming that $t=0$ is already in the late time regime, one does obtain}
 \[
 	R_c(t) = \begin{cases}
 	      \big( R_c(0)^3 + \tfrac{4}{9}\, t \big )^{1/3} &\quad \textrm{(DL)} \\
 	      \big( R_c(0)^2 + \tfrac{1}{2}\, t \big)^{1/2}             &\quad \textrm{(AL)}.
 	      \end{cases}
 \]
 For later purpose we note that this scaling law implies for $t,t_0 \gg R_c(0)^3$, that
 \begin{equation}
 \label{eq:t0}
 \frac{R_c(t_0)}{R_c(t)} = \Big( \frac{t_0}{t} \Big)^{1/\gamma}, \qquad
  \gamma = 3 \, \textrm{(DL)}, \quad  \textrm{or} \quad
  \gamma = 2    \, \textrm{(AL)}.
 \end{equation}
 Moreover, $z(\tau)$ may be obtained from \eqref{eq:vel_rescaled} by inverting the explicit solution of $\tau(z)$ given by
 \begin{align}
  \nonumber
\tau(z) & = \int \Big(  \nu\,\frac{z - 1}{z^\lambda} - z \Big)^{-1} \, dz \\
\label{eq:tau(z)}
        & =
\begin{cases}
 \frac{1}{2z - 3} -  \frac{4}{9} \ln (z+3) - \frac{5}{9} \ln(3 - z) & \quad \textrm{(DL)}\\
\frac{2}{z - 2} -  \ln (2-z) & \quad \textrm{(AL)}
\end{cases},
\end{align}
\new{where we have omitted a constant of integration, since later on only differences $\tau(z) - \tau(z_0)$ will be used.}
\new{Here the constant values of $\nu$ given in \eqref{eq:nu} for late times have been used.}
In Figure~\ref{fig:zTau} the solution $z(\tau)$ with initial value $z(0) = 5/4$ is depicted. Note that the reduced radius $z(\tau)$ is decreasing in time also for particles with $z > 1$, i.e. $R(t) > R_c(t)$. In fact, \new{one easily verifies, that} the right hand side of \eqref{eq:vel_rescaled} is negative for all particles with $z < 3/2$ (DL) or $z < 2$(AL), respectively.

The continuity equation~\eqref{eq:LSW1} for $f(z,\tau): = F\big(R_c z, t(\tau)\big) \, R_c$,
\begin{equation}
 \label{eq:LSW1_resc}
  \partial_\tau f + \partial_z \Big( \frac{dz}{d\tau}   f \Big) = 0,
\end{equation}
 can now be  solved by a separation ansatz $f(z,\tau) = g(\tau) h(z)$, yielding
 the scaled normalized island size distribution functions, \new{see \cite{Wagner1961},\cite{LifshitzSlyozov1961}}
\begin{align}
  \label{eq:hDL}
  h(z) & =
  \begin{cases}
  81 e 2^{-\frac{5}{3}}  z^2 (z+3)^{-\frac{7}{3}}(\tfrac{3}{2}-z)^{-\frac{11}{3}}\exp\big(\frac{-3}{3-2z}\big)
     &: \quad z \leq \frac{3}{2} \\
    0 &: \quad z > \frac{3}{2}
  \end{cases}  & & \text{(DL)}
\\
  \label{eq:hAL}
  h(z) & =
  \begin{cases}
  24 z(2-z)^{-5}\exp\big( \frac{-3z}{2-z} \big)
    &: \quad z \leq 2 \\
    0 &: \quad z > 2
  \end{cases}  & & \text{(AL)}
\end{align}
and the scaling
\begin{equation}
\label{eq:g}
g(\tau) = g_0 \exp(-3\tau) = g_0 \frac{R_c(0)^3}{R_c(t)^3}
\end{equation}
\new{Note that one outcome of the LSW analysis is a cutoff of the particle size distribution function with the largest rescaled radius
given by
\begin{equation}
\label{eq:zmax}
 z_\mathrm{max} = 3/2 \, \text{(DL),} \quad z_\mathrm{max} = 2\, \text{(AL)}.
\end{equation}
}
Finally we remark that the critical radius $R_c$ may be expressed in terms of the mean radius $\bar{R}$ as
$R_c = \bar{R}$ and $R_c = \tfrac{9}{8} \bar{R}$ for the  (DL) and the (AL) ripening, respectively \cite{Wagner1961}.


\section{Return radius}
\label{s:radius}
As has been sketched in the introduction, to calculate the newly formed volume between time $t_0$ and time $t$ we need to calculate what we have called
the {\it return radius} $r = r(t,t_0)$, i.e. the unique radius $r$ such that a particle with radius $R(t_0) = r$  will have the same radius at later time $t$, $R(t) = r$. In fact, this amounts to solve a boundary value problem for the differential equations \eqref{eq:vel_DL}, \eqref{eq:vel_AL}. We will see, however, that the return radius may be calculated quite easily in the asymptotic regime of LSW, i.e. for $t_0$ large enough, without explicitly solving the boundary value problem. Since in the following $t_0$ is fixed, we will write $r(t)$ instead of $r(t,t_0)$.

\begin{figure}[htbp]
\begin{center}	
	\includegraphics*[width=0.49\textwidth]{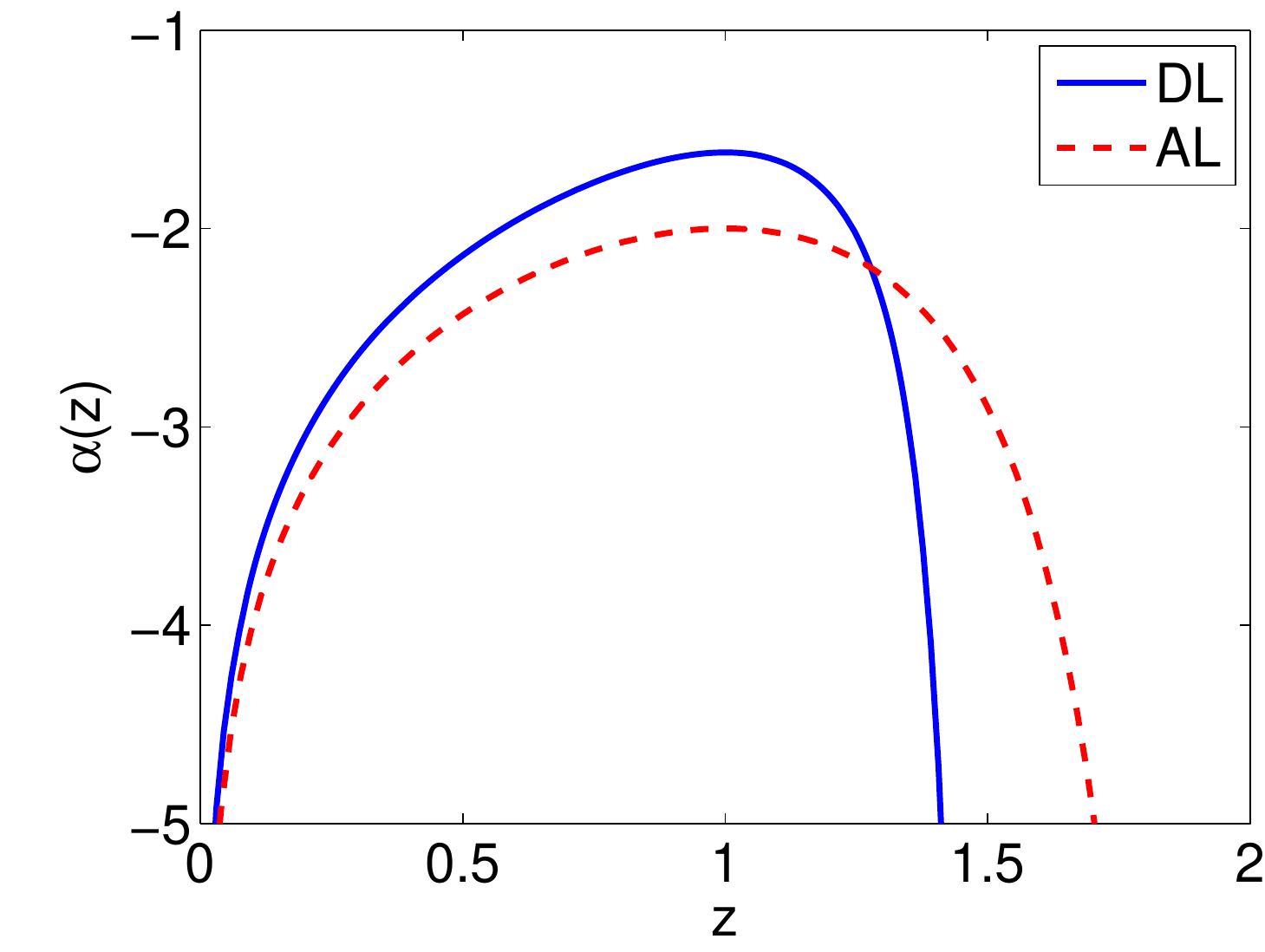}
		\includegraphics*[width=0.49\textwidth]{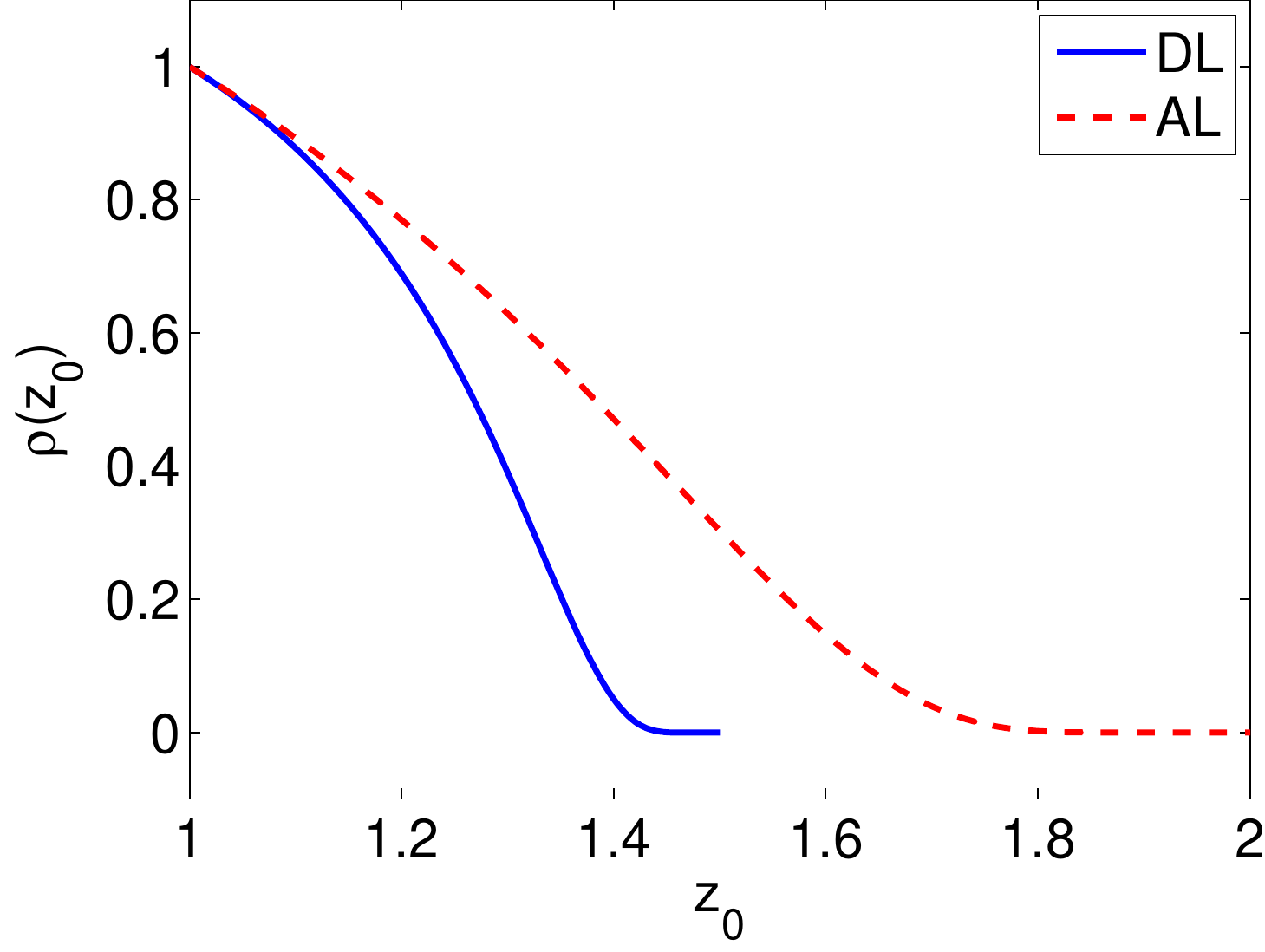}
	\caption{(Left) The function $\alpha(z) = \ln z + \tau(z)$ in \eqref{eq:aux2} has a unique maximum at $z = 1$. (Right) Choosing $z_0 > 1$, there is a unique $z:= \rho(z_0) < 1$ with $\alpha(z) = \alpha(z_0)$. The function $\rho$ has been evaluated numerically using bisection.}
	\label{fig:rhoZ}
	\end{center}
\end{figure}

We will use the rescaled coordinates $z,\tau$ as introduced in \eqref{eq:resc}. First note that in the rescaled coordinate $z$ the  return radius $r$ does take two different values at time $t_0$ and at time $t$, namely $z(t_0,r)$ and $z(t,r)$. We will use the notation
$\tau_0 = \tau(t_0)$, $\tau = \tau(t)$, $z_0 = z(\tau_0) = z(t_0,r)$, $z = z(\tau)=z(t,r)$. From  \eqref{eq:resc} follows, that $r$ being the return radius is equivalent to
\begin{equation}
\label{eq:aux1}
\ln z -  \ln z_0 = \tau_0  - \tau.
\end{equation}
Moreover, the scaling law \eqref{eq:t0} implies for large $t_0$
\begin{equation}
\label{eq:zScaling}
\frac{z}{z_0} = \frac{z(t,r)}{z(t_0,r)} = \frac{R_c(t_0)}{R_c(t)} = \Big( \frac{t_0}{t} \Big) ^\frac{1}{\gamma}.
\end{equation}
The two identities \eqref{eq:aux1} and \eqref{eq:zScaling} uniquely fix the return radius $r = r(t,t_0)$, which may be calculated as follows.
Since the time dependence of the rescaled radius $z = z(\tau)$ is given by equation \eqref{eq:vel_rescaled}, we use the explicit solution of $\tau(z)$ given in  \eqref{eq:tau(z)} to express \eqref{eq:aux1} as
\begin{equation}
\label{eq:aux2}
 \alpha(z) = \alpha(z_0), \quad \text{with} \quad \alpha(x) = \ln x + \tau(x) .
\end{equation}
Since the return radius has to be larger than the rescaled radius, $r \geq R_c(t_0)$, we may assume $z_0 = z(t_0,r) \geq 1$. One easily checks that
\[
\alpha^\prime(z)
\begin{cases}
	> 0 &: \quad z \in (0,1) \\
	= 0 &: \quad z = 1 \\
	< 0 &: \quad z \in (1,z_\mathrm{max})
\end{cases}
\]
and
 \[
 \lim_{z \to 0} \alpha(z) =\lim_{z \to z_\mathrm{max}} \alpha(z) =  -\infty.
\]
Thus, equation~\eqref{eq:aux2} allows for a unique solution for $z_0 \in [1,z_\mathrm{max})$.
 Denoting this inverse of $\alpha$ on $(-\infty,\alpha(1)]$ by $\psi$,  we may express $z = z(t,r)$ as a function $z = \rho(z_0)$
\begin{align}
\nonumber
\rho:  \, &[1,z_\mathrm{max}] \to [0,1] \\
\nonumber
\rho(z_0) &:=  \psi\big( \alpha ( z_0 ) \big), \quad z_0 \in (1,z_\mathrm{max}) \\
\label{eq:rho}
 \rho(1) &= 1, \, \rho(z_\mathrm{max}) = 0.
\end{align}
This function is easily evaluated numerically, e.g. using bisection, see Figure~\ref{fig:rhoZ}.

\begin{figure}[thbp]
\begin{center}	
	\includegraphics*[width=0.49\textwidth]{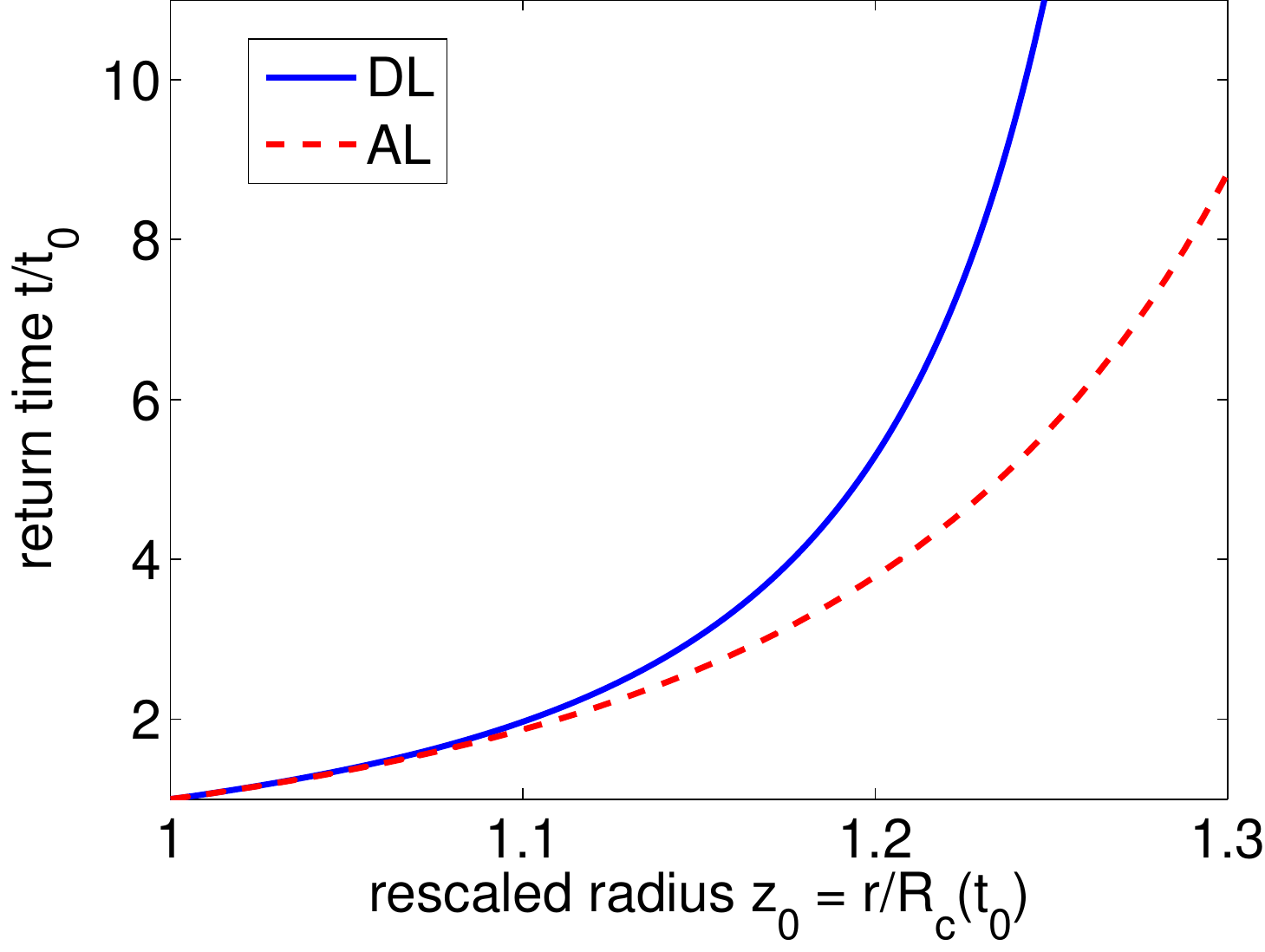}
		\includegraphics*[width=0.49\textwidth]{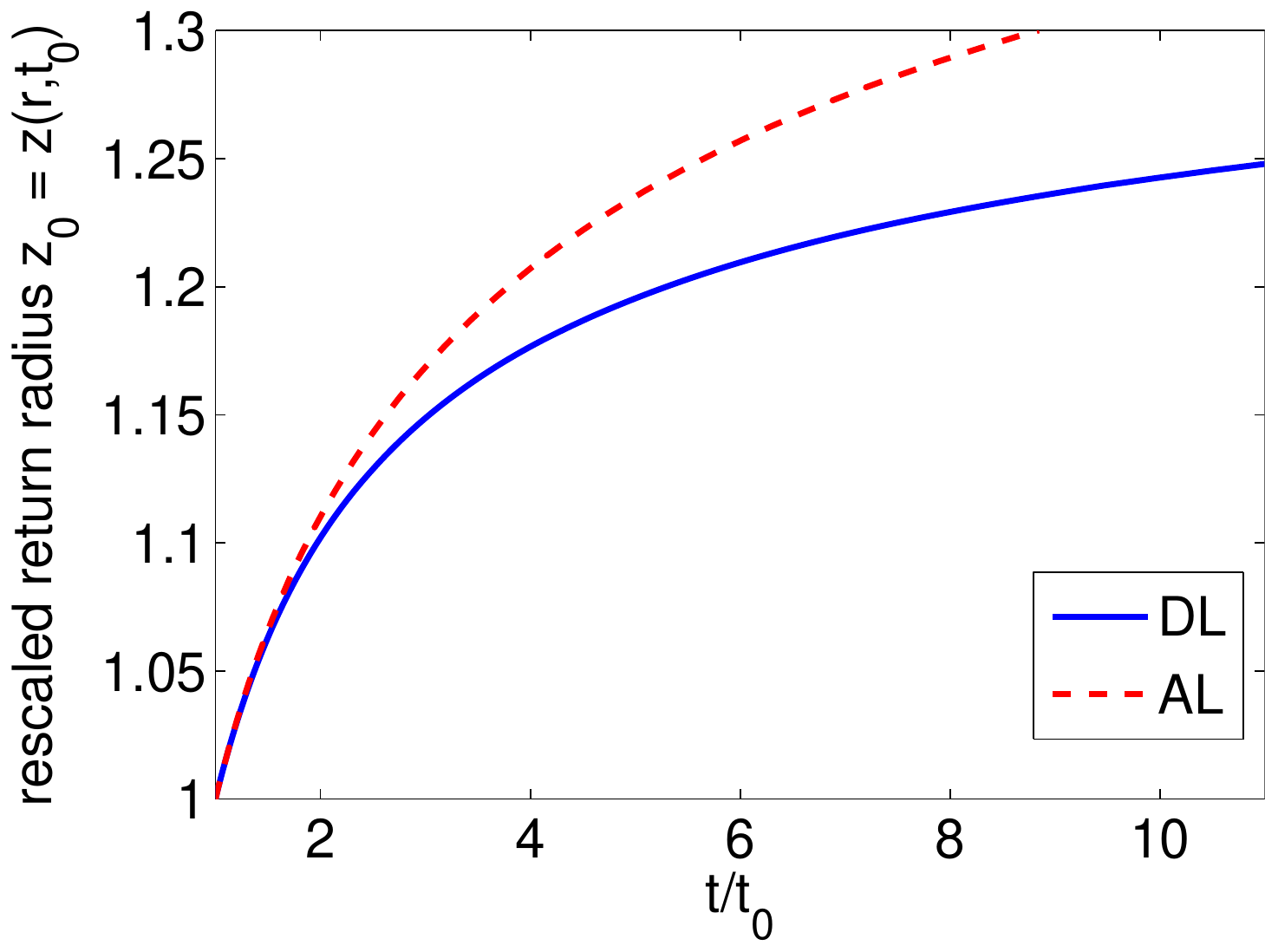}
	\caption{(Left) Return time over rescaled initial radius. (Right) Rescaled return radius over time. Time is given in units of the starting time $t_0$.}
	\label{fig:retRadius}
	\end{center}
\end{figure}

\begin{figure}[bhtp]
\begin{center}	
	\includegraphics*[width=0.9\textwidth]{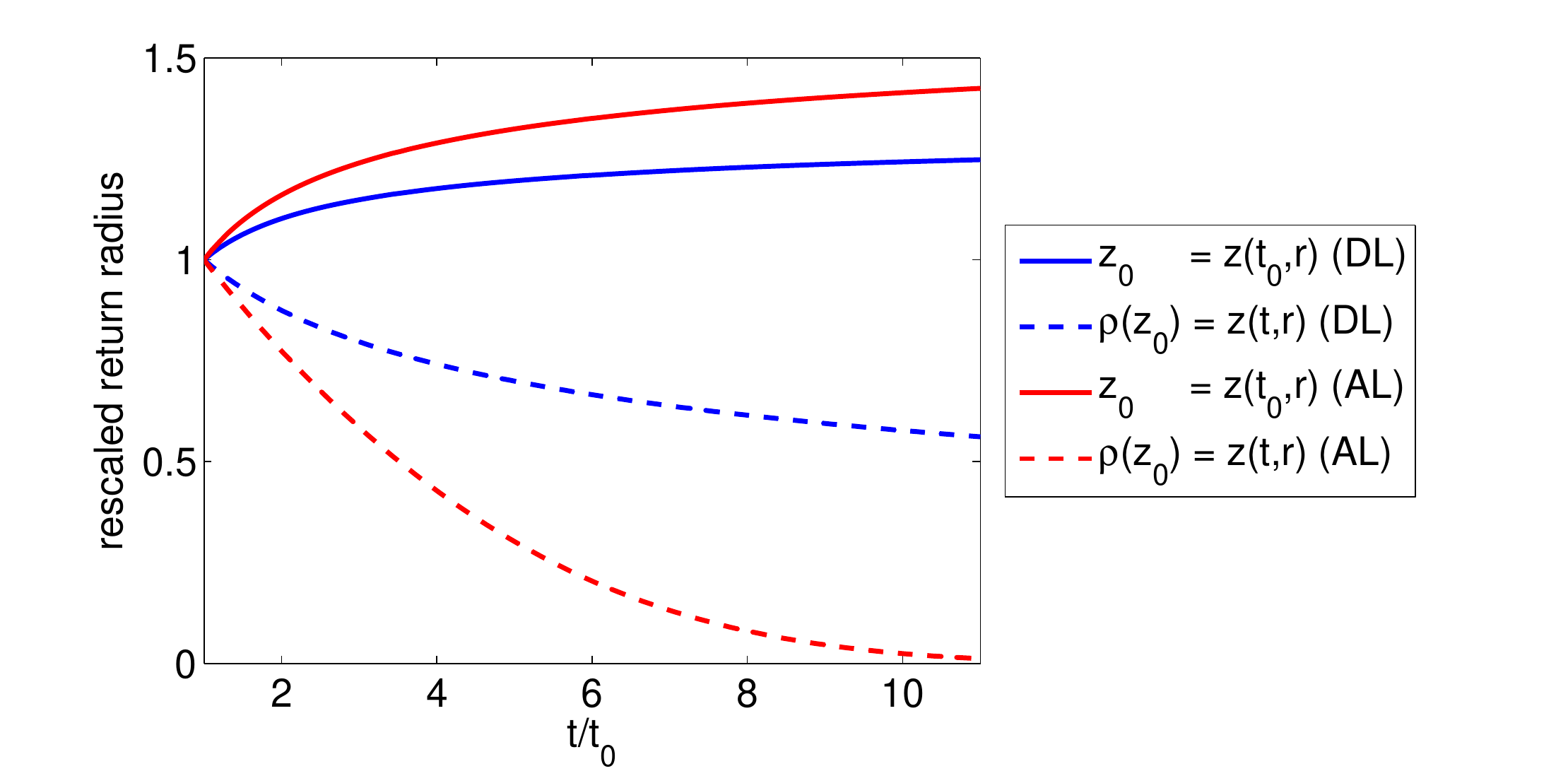}
		\caption{ The return radius $r = r(t,t_0)$ may be most easily calculated in rescaled coordinates using \eqref{eq:returnTime}.}
			\label{fig:z0rhoz0}
	\end{center}
\end{figure}

Now we may calculate the return time $t(r)$ for any initial radius $r$, i.e. the time $t$, such that $R(t) = R(t_0) = r$.
By equation~\eqref{eq:zScaling} we have
\begin{equation}
\label{eq:returnTime}
t(r) = t_0 \Big( \frac{z_0}{\rho(z_0)} \Big)^\gamma , \qquad z_0 = r / R_c(t_0)  .
\end{equation}
Inverting $t = t(r)$ in \eqref{eq:returnTime} numerically yields the return radius $r(t)$, see Figure~\ref{fig:retRadius}.
We point out, that according to \eqref{eq:returnTime}, the return radius $r(t,t_0)$ is a function  of $t/t_0$.
As it will turn out in the next section, we only need the pair $z_0, \rho(z_0)$ in order to calculate the amount of material which has been produced by recrystallization, see Figure~\ref{fig:z0rhoz0}.
Let us finally calculate the growth rate $\dot{r}(t)$ of the return radius at time $t = t_0$. Using \eqref{eq:returnTime}, we get
\[
 \dot{r}(t) = \Big( \frac {d}{dr} t(r) \Big)^{-1} = \Big( \frac {d t}{d z_0} \Big)^{-1} R_c(t_0).
\]
and equation \eqref{eq:returnTime} gives
\begin{equation}
\label{eq:aux4}
 \frac{dt}{dz_0} = t_0 \gamma \Big( \frac{z_0}{\rho(z_0)}\Big)^{\gamma - 1} \, \frac{\rho(z_0) - z_0 \rho^\prime(z_0)}{\rho(z_0)^2}.
\end{equation}
To proceed further, the derivative of $\rho(z_0)$ defined in \eqref{eq:rho} is needed at $z_0 = 1$. Since $\alpha^\prime(1) = 0$, $\alpha^{\prime\prime}(1) \neq 0$,  we have
\begin{equation}
\label{eq:aux5}
\rho(1 + h) = 1 - h + \mathcal{O} ( h^2 ),\quad \text{ i.e.} \,\,  \rho^\prime(1)  = -1.
\end{equation}
Evaluating \eqref{eq:aux4} at $z_0 = 1$, and using $\rho(1) = 1$ yields
\begin{equation}
\label{eq:dtdz}
	\frac{dt}{dz_0} \bigg|_{z_0 = 1} = 2 \gamma  t_0.
\end{equation}
So we end up with the growth rate
\begin{equation}
\label{eq:growthRadius}
 \dot{r}(t_0)  = \frac{R_c(t_0)}{2\gamma t_0}.
\end{equation}


\section{Volume fraction of recrystallized material}
\label{s:volume}
Now let $\Phi$ denote the volume fraction of the solid spherical phase, i.e.
\begin{equation}
\label{eq:Vol1}
 \Phi(t) = \tfrac{4}{3} \pi \int_0^\infty F(R,t) R^3 \, dR
\end{equation}
and denote by $\Phi^\mathrm{new}(t,t_0)$ the volume fraction of that part of the solid phase at time $t$ that was produced between time $t_0$ and time $t$ through recrystallization. Only those particles, which have grown between $t_0$ and $t$, i.e. $R(t) \geq R(t_0)$,
contribute. These particles are precisely the ones for which
 $R(t)$ is larger than the return radius $r = r(t,t_0)$, or, equivalently, $R(t_0) > r$. Thus $\Phi^\mathrm{new}(t,t_0)$ may be calculated as the difference of the volume fraction of all particles being larger than $r$ at time $t$ and the volume fraction of these same  particles at time $t_0$:
\begin{equation}
\label{eq:Vol2}
\Phi^\mathrm{new} (t,t_0) = \frac{4}{3} \pi \Big[
   \int_{r}^\infty F(R,t) R^3 \, dR -  \int_{r}^\infty F(R,t_0) R^3 \, dR   \Big].
\end{equation}
Assuming the LSW-theory to be valid, we may express $F(R,t)$ and $F(R,t_0)$ in terms of $f(z,\tau) = g(\tau) h(z)$ and
$f(z,\tau_0) = g(\tau_0) h(z)$, respectively, as given in \eqref{eq:hDL} - \eqref{eq:g}. Since $F(R,t) = R_c(t) f(z,\tau)$ and $F(R,t_0) = R_c(t_0) f(z,\tau_0)$, a change of variables in \eqref{eq:Vol2} leads to
\begin{equation}
\label{eq:Vol3}
\Phi^\mathrm{new} (t,t_0) = \tfrac{4}{3} \pi g_0 R_c(0)^3  \int_{z(t,r)} ^{z(t_0,r)} h(x) x^3\, dx
  = \tfrac{4}{3} \pi g_0 R_c(0)^3   \int_{\rho(z_0)} ^{z_0} h(x) x^3\, dx,
\end{equation}
where the return radius $r = r(t,t_0)$ may be calculated from equation~\eqref{eq:returnTime} as described in the last section.
In fact it is sufficient to solve for $z_0 = z(t_0,r)$ in order to evaluate~\eqref{eq:Vol3}, i.e. to invert the function $t = t(z_0)$
in \eqref{eq:returnTime} and to calculate $z(t,r) = \rho(z_0)$, see eq.~\eqref{eq:rho}.
Similarly we may express the total volume fraction in the late stage as
\begin{equation}
\label{eq:Vol4}
\Phi(t_0) = \Phi(t) = \tfrac{4}{3} \pi g_0 R_c(0)^3 \int_{0} ^{z_\mathrm{max}} h(x) x^3\, dx,
\end{equation}
Thus the percentage of the volume of  the second phase at time $t$, that has been produced by recrystallization
 between time $t_0$ and time $t$, is  given by
\begin{align}
\label{eq:VolFinal}
\varphi(t,t_0):= \Phi^\mathrm{new} (t,t_0) / \Phi
 =  \frac{1}{\overline{z^3}} \int_{\rho(z_0)} ^{z_0} h(x) x^3\, dx , \quad  \overline{z^3} := \int_{0} ^{z_\mathrm{max}} h(x) x^3\, dx,
\end{align}
and may be easily computed by determining $z_0 = z(t_0,r)$ and $\rho(z_0) = z(t,r)$ as described in section~\ref{s:radius} and using numerical quadrature. Again we point out, that $\varphi(t,t_0)$ may be expressed as a function of $t/t_0$, since $r(t,t_0)$ depends on $t/t_0$ only.

In Figure~\ref{fig:specVolume}, the specific produced volume $\varphi(t,t_0)$ is depicted over the normalized time $s:=t/t_0$.
As expected from \eqref{eq:VolFinal}, we observe that $\varphi(s)$ approaches the value $\varphi = 1$  for $s \to \infty$. Moreover, the growth of the recrystallized volume is nearly linear up to the value  $\varphi(s) \approx 1/4$.
 The growth rate of $\varphi(t,t_0)$ at time $t = t_0$ may be calculated using
equations~\eqref{eq:aux5}, \eqref{eq:growthRadius} as follows
\begin{align}
\nonumber
\overline{z^3} \, \frac{d}{dt} \varphi(t,t_0)\bigg|_{t=t_0} &= \frac{d}{d z_0} \int_{\rho(z_0)} ^{z_0} h(x) x^3\, dx \bigg|_{z_0 = 1} \,
\frac{d}{dt} z_0 (t)\bigg|_{t=t_0} \\
\nonumber
 &= \Big( h(z_0) - h(\rho(z_0)) \frac{d}{dz_0} \rho(z_0) \Big)\bigg|_{z_0 = 1}\,\frac{1}{2\gamma t_0} \\
 \label{eq:growthVolume1}
 & =  \frac{h(1)}{\gamma t_0}.
\end{align}
Here in the last equality we have again used that $\rho(1) = 1$ and $\rho^\prime(1) = -1$, see section~\ref{s:radius}.
 For the growth rate, we obtain from \eqref{eq:growthVolume1} and \eqref{eq:hDL}, \eqref{eq:hAL} the numerical values
\begin{equation}
\label{eq:growthVolume2}
\frac{d}{dt} \varphi(t,t_0)\bigg|_{t = t_0} =  \frac{h(1)}{\gamma t_0 \,  \overline{z^3}} =
\begin{cases}
27 e^{-2} \, 2^{-8/3} / ( \overline{z^3}\, t_0 ) \approx 0.51/t_0&: \quad \mathrm{DL} \\
12 e^{-3} / ( \overline{z^3} \, t_0 ) \approx    0.62 /t_0   &: \quad \mathrm{AL}
\end{cases}
\end{equation}

\section{Conclusions and outlook}
\label{s:outlook}
Let us summarize our findings:
We fix some time instant $t_0 \gg 1$ in the late stage of Ostwald ripening, where we assume that the LSW theory is valid.
All particles with radius $R(t_0)$ being larger than the critical radius $R_c(t_0)$, i.e. with renormalized radius
$$
 \qquad z_0 = r / R_c(t_0)  \in (1,z_\mathrm{max}]
$$
start to grow but -- except for the particle with maximum radius $z_\mathrm{max}$, \new{see \eqref{eq:zmax}} -- will start to shrink at some later time instant.
For every such $z_0$ we can find the return time $t(z_0)$, i.e. the time instant at which
the particle will return to its initial size, from the monotonic formula (see \eqref{eq:returnTime})
$$
t(z_0) = t_0 \, \left ( \frac{z_0}{\rho(z_0)} \right )^\gamma.
$$
The function $\rho(z)$  is defined in  eq.~\eqref{eq:rho} and its graph is given in Figure~\ref{fig:rhoZ}.
The inverse function $z_0(t)$ is shown in Figure~\ref{fig:retRadius}.

\begin{figure}[bhtp]
\begin{center}	
	\includegraphics*[width=0.7\textwidth]{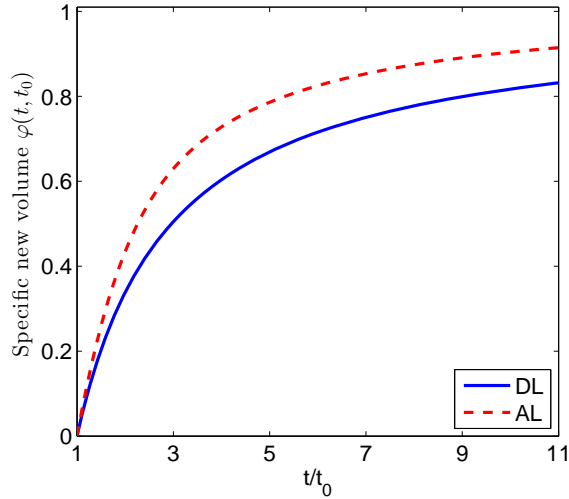}
	\caption{ Specific volume $\varphi(t,t_0)$ of recrystallized material from time $t_0$ to time $t$ over normalized time $t/t_0$ in the late stage, where LSW theory is valid. $\varphi(t,t_0)$ is computed numerically by first solving $(t/t_0)^{1/\gamma} = z_0/\rho(z_0)$  for $z_0 = z(r,t_0)$, with $\rho$ given in \eqref{eq:rho}, and then calculating the integral $\int_{\rho(z_0)} ^{z_0} h(x) x^3\, dx$  using numerical quadrature.
The slope at $t/t_0 = 1$ is given in \eqref{eq:growthVolume2}.}
	\label{fig:specVolume}
	\end{center}
\end{figure}

A formula for the specific volume $\varphi(t,t_0)$ of the newly formed material within the time interval $[t_0,t]$ has been derived for both, $AL$ and $DL$ ripening. It is given in terms of the corresponding scaled normalized island size distribution function $h(z)$ as
\begin{align}
\label{eq:VolFinal2}
\varphi(t,t_0)= \frac{1}{\mbox{const}} \int_{\rho(z_0(t))} ^{z_0(t)} h(x) x^3\, dx .
\end{align}
The graph of $\varphi$ is given in Figure~\ref{fig:specVolume}.
Moreover, an explicit formula for the growth rate $\frac{d}{dt}\varphi(t,t_0)|_{t = t_0}$ is given in
\eqref{eq:growthVolume2}.

In a forthcoming paper we plan to apply our findings to the
isotope uptake during Ostwald ripening. Here one also has to account for the adsorption of the tracer material at the surface of the solid phase and therefore the change of surface area during ripening may play a role. Moreover, a generalization to the case of large volume fractions, where the LSW theory is not valid, needs further investigations.

\bibliographystyle{plain}

\bibliography{OstwaldMath}

\end{document}